\documentclass[12pt,letterpaper]{article}
\usepackage[utf8]{inputenc}
\usepackage[english]{babel}
\usepackage[vmargin=0.6in,hmargin={1in,1in},headsep=0.3in,headheight=1in]{geometry}
\usepackage{subfiles}
\usepackage{tikz}
\usepackage{fix-cm}
\usepackage{enumitem}
\usepackage{multicol}
\usepackage{amsmath, amsthm, amssymb, amsfonts}
\allowdisplaybreaks
\newcounter{taggedEquations}
\let\OldTag\tag
\renewcommand*{\tag}[1]{\stepcounter{taggedEquations}\OldTag{#1}}
\usepackage{array}
\usepackage{relsize}
\usepackage{xcolor}
\usepackage{lastpage}
\usepackage{blindtext}
\usepackage[numbers]{natbib}
\usepackage[linktocpage=true]{hyperref}
\hypersetup{
	colorlinks,
	citecolor=blue,
	filecolor=black,
	linkcolor=blue,
	urlcolor=blue
}
\usepackage{url}
\newcommand{\overbar}[1]{\mkern 2.5mu\overline{\mkern-2.5mu#1\mkern-2.5mu}\mkern 2.5mu}
\usepackage{graphicx}
\usepackage{subcaption}
\usepackage{wrapfig}
\usepackage{indentfirst}
\usepackage{hyperref}
\usepackage{soul}
\usepackage{empheq}
\usepackage{authblk}
\setlist{parsep=0pt,listparindent=\parindent}
\setlist[itemize]{noitemsep, topsep=0pt}
\setlist[enumerate]{noitemsep, topsep=0pt}
\setlist{parsep=0pt,listparindent=\parindent}

\stepcounter{footnote}
\graphicspath{{Images/}{../Images/}}
\begin{document}  
\def\UrlFont{}
\title{\Large\textbf{An Invariant Characterization of the Levi-Civita Spacetimes}}
\author[1,2]{{\large{C.~K.~Watson\footnote{mail to: \em\texttt{\href{cooper_watson@Baylor.edu}{Cooper\_Watson@Baylor.edu}}}}}}
\author[1,2]{\large{W.~Julius}}
\author[1,2]{\large{M.~Gorban}}
\author[3]{\large{D.~D.~McNutt}}
\author[1]{\large{E.~W.~Davis}}
\author[1,2]{\large{G.~B.~Cleaver}}

\affil[1]{\emph{Early Universe, Cosmology and Strings (EUCOS) Group, Center for Astrophysics, Space Physics and Engineering Research (CASPER), Baylor University, Waco, TX 76798, USA} \vspace{0.25cm}}

\affil[2]{\emph{Department of Physics, Baylor University,  Waco, TX 76798, USA} \vspace{0.25cm}}

\affil[3]{\emph{Faculty of Science and Technology, University of Stavanger, N-4036 Stavanger, Norway}}

\date{\today}  
\maketitle  
\begin{abstract}   
In the years 1917--1919 Tullio Levi-Civita published a number of papers presenting new solutions to Einstein's equations. This work, while partially translated, remains largely  inaccessible to English speaking researchers. In this paper we review these solutions, and present them in a modern readable manner. We will also compute both Cartan--Karlhede and Carminati--Mclenaghan invariants such that these solutions are invariantly characterized by two distinct methods. These methods will allow for these solutions to be totally and invariantly characterized. Because of the variety of solutions considered here, this paper will also be a useful reference for those seeking to learn to apply the Cartan--Karlhede algorithm in practice.
\end{abstract} 

\par

\section{Introduction}
In the years 1917--1919 Tullio Levi-Civita (LC) published nearly a dozen papers introducing and analyzing a variety of new solutions to Einstein's field equations (collected works in Italian available in Volume IV at \cite{collectedLC}). Recently, several key papers have been republished in English, including two of Levi-Civita's original papers \cite{levirepublication1917,levirepublication1919}, and \cite{jordan2009republication} containing an overview of several solutions not included in any of the other translations. In \cite{levirepublication1917}, a homogeneous Einstein--Maxwell spacetime is derived; in \cite{levirepublication1919}, a spacetime with a potential analogous to the logarithmic Newtonian gravitational potential is derived; and \cite{jordan2009republication} discusses derivation of several degenerate vacuum spacetimes. There are additional spacetime solutions in literature not translated into English \cite{levi1918ds2quadrantali,levi1918ds2oblique} which are similar, but distinct from the other degenerate vacuum solutions. 

The age and structure of these papers has resulted in more contemporary works citing these papers in confusing or incorrect ways. Here, we clarify the structure of works on exact solutions that are of interest. First, a homogeneous solution was published in 1917 as a standalone paper \cite{levi1917italian}. Then, in the years from 1917 to 1919, a series of nine notes were published starting with \cite{levi1917ds2start} and ending with \cite{levi1918ds2log}. It is not uncommon to find the different papers in this series cited by the general heading of the entire series or by referencing to only the first article in the series. 

Additionally, we will provide an invariant (local) characterization of these solutions via two different methods. First, we will utilize the Cartan--Karlhede (CK) algorithm \cite{karlhede1980review} to generate an invariant coframe and the corresponding scalar quantities which uniquely characterize these spacetimes. The variety of solutions considered in this work will result in the CK algorithm running in several markedly different ways. We will present an overview of the algorithm itself, as well as a comprehensive guide that fully outlines and computes each step of the CK algorithm for the different spacetimes. Thus, this paper should serve as a useful resource for those attempting to learn to apply the CK algorithm in practice. 

Second, we calculate the Carminati--McLenaghan (CM) \cite{carminati1991algebraic} scalar invariants to construct a coframe independent classification of each solution. The set of CM invariants is advantageous as they are of the lowest possible degree and are generally the minimal independent set for any valid Petrov and Segre type spacetime. All spacetimes considered herein are of a Segre type, such that only CM invariants are needed, i.e., we do not need the extended set of invariants given in \cite{zakhary1997complete}. In fact, in several of the cases considered the spacetime is sufficiently special, such that only a subset of the CM invariants are needed~\cite{santosuosso}. We do note explicitly that the CM invariants will only uniquely characterize these solutions to zeroth order (in derivatives), but such invariants are useful for distinguishing LC solutions. In several cases, we will also present ``$\mathcal{I}$'' invariants \cite{abdelqader2015invariant}, as these invariants are distinct from the CM invariants and may contain information regarding algebraically special surfaces \cite{abdelqader2015invariant, mcnutt2021geometric}. For completeness, we note that all spacetimes considered here are $\mathcal{I}$ non-degenerate, as the only case considered with constant scalar invariants is homogeneous \cite{coley2009spacetimes}. We also note that the CK algorithm will always generate a complete classification of the spacetime, thus there is no possibility this may fail for the specific cases considered here.

We will also consider several generalizations of these solutions in cases where our methods of characterization extend directly, and in an instructive manner. In particular, we are interested in the work given by \cite{bertotti1959uniform}, which generalizes the solution in \cite{levirepublication1917,levi1917italian} to one which is not conformally flat, and in \cite{delice2004kasner} which generalizes the work in \cite{levirepublication1919,levi1918ds2log} to a solution which is not generally static.

Throughout, the $(-1,1,1,1)$ signature convention will be used. Greek indices will be taken to run from $0$ to $3$, and follow the Einstein summation convention. Parentheses (brackets) around indices will denote the usual (anti)-symmetrized indices. Partial derivatives in the form $\frac{\partial}{\partial x}$ will always be taken to be covariant, and null tetrads will always be listed in covariant form. The null vectors $\{l,k,m,\overbar{m}\}$ will be taken to have normalization, such that $l_\mu k^\mu=-1$ and $\overbar{m}_\nu m^\nu=1$.

\section{Overview of the CK Algorithm}

Here, we provide an overview of the practical CK algorithm used throughout this paper. For a review of the theoretic underpinning of the general Cartan process, see \cite{olver1995equivalence}. For a review of this process's application to general relativity, refer to \cite{karlhede1980review,stephani2009exact,maccallum1994algebraic}.
In this algorithm we will use $q$ to denote the order of differentiation, which tracks the current iteration of the algorithm. In the steps given here, we will also depart from the ``standard'' description of the algorithm by treating the $q=0$ order step as a distinct ``initialization" step and all steps with $q\geq1$ as the repeated part of the algorithm. We do this as the zeroth order step is the only step in which we will require full knowledge of the algebraic type of the tensors considered as it will usually be the step at which the parameters of the isotropy group are fixed the most.

 The CK algorithm will run as follows:
\begin{enumerate}
    \item Take the order of differentiation to be $q=0$;\\
\vspace{-12pt}
    \item Determine the Petrov and Segre types of the spacetime. Practical algorithms for this can be found in \cite{zakhary2003new,zakhary2004new}, respectively. These types will be used to determine the possible invariant forms to use at zeroth order;\\
    \vspace{-12pt}
    \item Construct a null tetrad for the spacetime;\\
    \vspace{-12pt}
    \item Calculate the components of the Riemann tensor along the current null tetrad. It will be useful to split the Riemann tensor into its irreducible parts;\\
    \vspace{-12pt}
    \item Using the known Petrov and Segre types, along with the forms of the curvature computed above, determine an invariantly defined frame for the spacetime which fixes the frame as much as possible. Here we will make use of the invariant forms given in \cite{pollneystandardforms}. We also note that while it is possible to start with any frame and determine the transform which brings it to its standard form, we will usually try to determine a frame which is in (or as close as possible to) an invariant form at zeroth order. At this step, one will often have to select, by hand, if one is setting the Ricci or Weyl tensor into an invariant form, as it is not generally possible to find a frame which fixes both tensors into their canonical form; \\
    \vspace{-12pt}
    \item Using this canonical form, determine the number of functionally independent terms which are now invariantly defined by the given frame. One method of doing this is constructing the Jacobian for the functions and determining its rank;\\
    \vspace{-12pt}
    \item Set $q=1$;\\
    \vspace{-12pt}
    \item Calculate the $q$th derivatives of the tensor which has been set into an invariant form;\\
    \vspace{-12pt}
    \item Determine the isotropy group which leaves these derivatives invariant. This group will be a subgroup of the isotropy group at order $q-1$, and thus one only needs to check how the $q$th derivatives transform under the $(q-1)$st isotropy group and find the new maximal invariant subgroup;\\
    \vspace{-12pt}
    \item If the new invariant subgroup is smaller than the previous group, fix the transformation parameters such that the derivatives are in an invariant form;\\
    \vspace{-12pt}
    \item Determine the number of new functionally independent terms appearing at order $q$;\\
    \vspace{-12pt}
    \item If the isotropy group and number of functionally independent terms has not changed from the $q-1$ step, the algorithm terminates. The full set of CK invariants are all of the derivative components computed thus far. If the isotropy group or functionally independent terms has changed, then set $q=q+1$ and return to step 8.
\end{enumerate}

\section{The Homogeneous Levi-Civita Solution (1917)}
In 1917, Levi-Civita presented a solution to Einstein's field equations which described a space permeated by a homogeneous, non-null Maxwell field \cite{levirepublication1917,levi1917italian}. This solution was later rediscovered independently (and nearly simultaneously) in \cite{bertotti1959uniform,robinson1959solution}, and, as such, is often called the Bertotti--Robinson metric in literature. It was \cite{bertotti1959uniform} that presented a slight generalization (discussed below) which has a non-vanishing cosmological constant. It was shown in \cite{dolan1968singularity} that this metric was generally singularity free. This spacetime can also be shown to be a limiting case of the more general Petrov solution \cite{bonnor1979source}.

Later considerations of more general Maxwell spacetimes have also revealed several interesting properties regarding this solution. First, this is the only homogeneous non-null Maxwell solution \cite{kramer1978homogeneous}. Generalizations where the Maxwell field does not share the homogeneous symmetry also give interesting solutions (not discussed here) which are algebraically more general \cite{mclenaghan1975new,tariqtupper1975class}. 

\subsection{Forms of the Metric and Nature of the Coordinates}

In present literature, there are at least six equivalent forms of the homogeneous 1917 Levi-Civita line element. Here, we will introduce these different forms and discuss the relations between them when possible and mention any relevant coordinate artifacts that might be present. We will also present several new forms of this metric and discuss several cases in which no coordinate transforms exist in the literature, nor can  it be derived without the use of complex transforms.

The original form of the line element \cite{levirepublication1917} was given in cylindrical coordinates as:
\begin{equation}
ds^2=-\left(c_1e^{z/a}+c_2e^{-z/a}\right)^2dt^2+d\rho^2+a^2\sin\left(\rho/a\right)^2d\phi^2+dz^2 , \label{met 1917}
\end{equation}
\noindent where $c_1$ and $c_2$ are integration constants originally derived by Levi-Civita and $a$ is a constant usually associated with an electric or magnetic field strength \cite{levirepublication1917,levi1917italian}. If $c_1$ and $c_2$ are both non-zero and of opposite sign, this metric will be singular for $z=\frac{a}{2}\ln{\lvert\frac{c_2}{c_1}\rvert}$. Additionally, this metric is singular for $\rho=n \pi a$, where $n\in\mathbb{Z}$.

One may define an angular parameter $\theta=\rho/a$, such that \eqref{met 1917} can be rewritten in the form given in \cite{puthoff} as: 
\begin{equation}
ds^2=-\left(c_1e^{z/a}+c_2e^{-z/a}\right)^2dt^2+dz^2+a^2d\Omega^2, \label{met puthoff}
\end{equation}

\noindent where $d\Omega^2$ is the line element for the 2-sphere. We explicitly note that $\theta\in\mathbb{R}$. This metric can also be be put into a secondary set of cylindrical coordinates \cite{puthoff} by taking $\rho^\prime/a=\sin{\theta}$ giving: 
\begin{equation}
ds^2=-\left(c_1e^{z/a}+c_2e^{-z/a}\right)^2dt^2+\frac{d\rho^{\prime\,2}}{\left(1-\left(\rho'/a\right)^2\right)}+\rho^{\prime\,2}d\phi^2+dz^2, \label{met1}
\end{equation}

 We note here that $\rho^\prime\in\left[0,a\right]$. In these coordinates the countably infinite singularities in the original radial coordinate have been reduced to only $\rho=0$ and $\rho=a$.

We may once again rewrite \eqref{met 1917}, this time in the associated Cartesian coordinates, previously exploited in \cite{paulibook}, as:
\begin{equation}
ds^2=-\left(c_1e^{z/a}+c_2e^{-z/a}\right)^2dt^2+dx{'\,^2}+dy{'\,^2}+dz^2+\frac{\left(x'\,dx'+y'\,dy'\right)^2}{a^2-\left(x{'\,^2}+y{'\,^2}\right) },\label{met pauli}
\end{equation}

\noindent where we explicitly note that the primed coordinates $x'$ and $y'$ are restricted to take values subject to $x{'\,^2}+y{'\,^2}\leq a^2$. 

It is also possible to eliminate $c_1$ and $c_2$ via a transform on both $t$ and $z$. By scaling $t$ and translating $z$, the original metric can be rewritten into one of two equivalent forms:
\begin{gather}
ds^2=-\sinh(z)^2dt^2+a^2dz^2+a^2d\Omega^2\\
\text{or} \notag\\
ds^2=-\cosh(z)^2dt^2+a^2dz^2+a^2d\Omega^2,
\label{met stephani 1}
\end{gather} 

\noindent where, for real coordinate transforms, one can get either a $\emph{sinh}$ or $\emph{cosh}$ solution depending on the relative signs of $c_1$ and $c_2$. Interestingly, it appears that the only method of connecting these two \textit{equivalent} solutions is via complex transformations on both $t$ and $z$, although it is not presently understood why this method works. We do note that this property of certain complex transformations, resulting in the same solution in different forms, is remarkably similar to the application of the Newman--Janis trick when applied to Minkowski space~\cite{rajan2016complex}.

Here we also present a new form of this metric which is not related to previous solutions by any known transform, real or complex: 
\begin{equation}
    ds^2=-e^{2z}dt^2+a^2dz^2+a^2d\Omega^2.
\end{equation}

Using this form and making the coordinate transform $r=e^{-z}$ (along with a rescaling in $t$) we get the form seen in \cite{stephani2009exact}, given as:
\begin{equation}
    ds^2=\frac{a^2}{r^2}\left(dr^2-dt^2\right)+a^2d\Omega^2. \label{primary}
\end{equation}

This reference also presents the form
\begin{equation}
    ds^2=-\left(1+\frac{z^2}{a^2}\right)dt^2+\left(1-\frac{y^2}{a^2}\right)dx^2+\left(1-\frac{y^2}{a^2}\right)^{-1}dy^2+\left(1+\frac{z^2}{a^2}\right)^{-1}dz^2,\label{reduction}
\end{equation}
\noindent which we see is a special case of the more general metric considered in \eqref{bertotti}.

Throughout the rest of this paper we will work with the metric and coordinates given by \eqref{primary}. A convenient choice of covariant null tetrad is given by:

\begin{equation}
\begin{gathered}
    l=\frac{a}{\sqrt{2}r}\left(\frac{\partial}{\partial r}-\frac{\partial}{\partial t}\right), \quad
    k=-\frac{a}{\sqrt{2}r}\left(\frac{\partial}{\partial r}+\frac{\partial}{\partial t}\right), \\
    m=\frac{a}{\sqrt{2}}\left(\frac{\partial}{\partial \theta}+i\sin{\left(\theta\right)}\frac{\partial}{\partial \phi}\right), \quad
    \overbar{m}=\frac{a}{\sqrt{2}}\left(\frac{\partial}{\partial \theta}-i\sin{\left(\theta\right)}\frac{\partial}{\partial \phi}\right).
\end{gathered}\label{null}
\end{equation}

\subsection{Curvature Invariants and CK Classification}

Using the null tetrad defined in \eqref{null}, one may calculate the only non-vanishing curvature component
\begin{equation}\label{cart}
    \Phi_{00}=\frac{1}{2a^2}.
\end{equation}

 This constant solution is therefore both conformally flat and Ricci flat. 
Since this curvature component is constant and is an invariantly defined frame, \eqref{cart} is the only non-vanishing CK invariant \cite{pollneystandardforms}.

Additionally, this result fixes all CM invariants to be either zero or of the form $c a^{2n}$ for $a,c\in\mathbb{R} \text{ and } n\in\mathbb{Z}$. For example, the only two non-vanishing CM invariants are
\begin{equation}
    r_1=2\sqrt{r_3}=\frac{1}{a^4},
\end{equation}
\noindent where only $r_1$ is independent since the spacetime is of warped product type $B_2$ \cite{santosuosso}.

\subsection{Regarding Electromagnetic ``Wormholes"} 

It has been suggested that this solution may in some sense constitute a wormhole supported by electromagnetic stress \cite{maccone1995interstellar,maccone2000seti}. It was shown in \cite{puthoff,davis1999wormhole,davis1998interstellar} that, despite there being a coordinate singularity at $\rho'=a$ in \eqref{met1}, this does not correspond to a wormhole throat, as an appropriate choice of coordinate transform can be made, such that the spatial part of the metric becomes that of a hypercylinder. Here, we will highlight two different methods of characterizing this surface, which provide a secondary method of determining that this solution is not a wormhole. 

Working with the coordinates and metric given by \eqref{met1}, we choose the following null tetrad to analyze the surface $\rho=a$, where we have dropped the prime out of convenience,
\begin{equation}
\begin{split}
    l&=\frac{1}{\sqrt{2}}\left(-\left|c_1e^{z/a}+c_2e^{-z/a}\right|\frac{\partial}{\partial t}+\frac{a}{\sqrt{a^2-\rho^2}}\frac{\partial}{\partial \rho}\right), \\
    k&=\frac{1}{\sqrt{2}}\left(-\left|c_1e^{z/a}+c_2e^{-z/a}\right|\frac{\partial}{\partial t}-\frac{a}{\sqrt{a^2-\rho^2}}\frac{\partial}{\partial \rho}\right), \\
    m&=\frac{1}{\sqrt{2}}\left(\rho\frac{\partial}{\partial \phi}+i\frac{\partial}{\partial z}\right), \quad \overbar{m}=\frac{1}{\sqrt{2}}\left(\rho\frac{\partial}{\partial \phi}-i\frac{\partial}{\partial z}\right).
\end{split}
\end{equation}

 The null expansions along $l$ and $k$ are, respectively,
\begin{equation}
\theta_{(l)} = \overbar{q}^{\,\mu\nu} \nabla_\mu l_\nu~ ~~~\text{and}~~~ \theta_{(k)} = \overbar{q}^{\,\mu\nu} \nabla_\mu k_\nu~,
\end{equation}
\noindent where $\overbar{q}_{\mu\nu} = g_{\mu\nu} +2 l_{(\mu} n_{\nu)}$ is a local, induced two-metric and $\nabla_\mu$ is the standard covariant derivative. In terms of the given null frame, these are explicitly:
\begin{equation}
    \theta_{(l)}=-\theta_{(n)} =-\frac{\sqrt{a^2-\rho^2}}{\sqrt{2}a\rho}.
\end{equation}

 It can be seen that the surface $\rho=a$ is indeed a surface on which the expansion of these null directions vanish. For this surface to correspond to a wormhole throat, the derivative of the expansion projected along the null direction must be positive on the surface \cite{hvwormholes1998}. Computing these terms at the surface we can show that
\begin{equation}
    l^\mu\nabla_\mu\theta_{(l)}=k^\mu\nabla_\mu\theta_{(k)}=-\frac{1}{2a^2}.
\end{equation}

Thus, this surface cannot correspond to a wormhole throat. It is in fact only a maximal surface relating to the given range of coordinates \cite{landis1997magnetic,senovilla2013trapped}. This surface is also not detectable by any scalar invariants (since all invariants are constant) and, thus, is not a geometric surface, as would be expected for a wormhole throat \cite{mcnutt2021geometric}.

\subsection{Bertotti Generalization}

In \cite{bertotti1959uniform}, a generalization of the 1917 solution is used, with the line element taking the form of
\begin{equation}
    ds^2=-\left(1+\frac{y^2}{r_+^2}\right)dt^2+\left(1-\frac{z^2}{r_-^2}\right)dx^2+\left(1+\frac{y^2}{r_+^2}\right)^{-1}dy^2+\left(1-\frac{z^2}{r_-^2}\right)^{-1}dz^2~, \label{bertotti}
\end{equation}
\noindent where $x,y,z$ are Cartesian coordinates. If $r_+\neq r_-$, this solution is no longer conformally flat or Ricci flat. In the case where $r_+ = r_-=a$, the spacetime becomes conformally flat and reduces to \eqref{reduction}. 

Using the null tetrad:
\vspace{-8pt}
\begin{equation}
\begin{split}
    l&=\frac{1}{\sqrt{2}}\left(-\sqrt{1+\frac{x^2}{r_+^2}}\frac{\partial}{\partial t}+\frac{1}{\sqrt{1+\frac{x^2}{r_+^2}}}\frac{\partial}{\partial x}\right), \quad
    k=\frac{1}{\sqrt{2}}\left(-\sqrt{1+\frac{x^2}{r_+^2}}\frac{\partial}{\partial t}-\frac{1}{\sqrt{1+\frac{x^2}{r_+^2}}}\frac{\partial}{\partial x}\right),\\
    m&=\frac{1}{\sqrt{2}}\left(\sqrt{1-\frac{z^2}{r_-^2}}\frac{\partial}{\partial y}+\frac{i}{\sqrt{1-\frac{z^2}{r_-^2}}}\frac{\partial}{\partial z}\right),\quad 
    \overbar{m}=\frac{1}{\sqrt{2}}\left(\sqrt{1-\frac{z^2}{r_-^2}}\frac{\partial}{\partial y}-\frac{i}{\sqrt{1-\frac{z^2}{r_-^2}}}\frac{\partial}{\partial z}\right),
\end{split}
\end{equation}
\noindent the only non-vanishing curvature components are found to be
\begin{equation}
\begin{gathered}
R=2\left(\frac{1}{r_-^2}-\frac{1}{r_+^2}\right), \\ \Phi_{11}=\frac{1}{4}\left(\frac{1}{r_+^2}+\frac{1}{r_-^2}\right), \quad \Psi_2=\frac{1}{6}\left(\frac{1}{r_+^2}-\frac{1}{r_-^2}\right), \label{bertottick}
\end{gathered}
\end{equation}
\noindent which are again all constants. Since both terms in \eqref{bertottick} are constant, and this is already an invariantly defined frame to zeroth order, these are the only three non-vanishing Cartan invariants.

Since this spacetime is a warped product spacetime of type $B_2$, the four invariants given below are the complete set \cite{santosuosso}. These are:
\begin{equation}
    R=2\left(\frac{1}{r_-^2}-\frac{1}{r_+^2}\right), \quad r_1= \frac{1}{4}\left(\frac{1}{r_-^2}+\frac{1}{r_+^2}\right)^2, \quad r_2=0, \quad w_2=\frac{1}{36}\left(\frac{1}{r_-^2}-\frac{1}{r_+^2}\right)^3.
\end{equation}

\section{The Cylindrical Levi-Civita Solution (1919)}
In \cite{levirepublication1919,levi1918ds2log}, a solution to Einstein's field equations is presented that serves as the analog to the Newtonian logarithmic potential. In \cite{silva1995santos}, the CK invariants have already been computed, and in \cite{herrera1999levi} these invariants were used to show that this solution can be found as a limiting subcase of the $\gamma$ (or ``Zipoy-Voorhees'') \cite{zipoy1966topology,voorhees1970static} solution. Here, for completeness, we will independently compute the CK invariants. 

This spacetime solution has also been of interest as an exterior vacuum solution to various physical sources, see \cite{silva1995santos,wangcylindrical}. There has also been interest in global (topological) properties of this solution as certain parameters (which do no affect local properties) are related to cosmic strings, see \cite{wangcylindrical,wang1997parametersCOSMICSTRINGS,krisch2011levi} for discussion of these properties as they pertain to this solution, and see \cite{hindmarsh1995cosmic} for a general review of cosmic strings.
\subsection{Forms of the Metric and Nature of the Coordinates}

The line element for this spacetime is given by
\begin{equation}
    ds^2=-e^{2\nu}dt^2+e^{-2\nu}\left(e^{2\lambda}\left(d\rho^2+dz^2\right)+r^2d{\phi}^2\right), \label{1919 LC metric}
\end{equation}
\noindent in standard cylindrical coordinates, where $\nu(\rho,z)$ is a solution to
\begin{equation}
    \frac{1}{\rho}\frac{\partial}{\partial \rho}\left(\rho\frac{\partial{\nu}}{\partial{\rho}}\right)+\frac{\partial^2\nu}{\partial{z}^2}=0,
\end{equation}
\noindent and $\lambda$ is defined, up to a constant of integration, by the differential relation:
\begin{equation}
    d\lambda=\rho\left(\frac{\partial{\nu}}{\partial{\rho}}^2-\frac{\partial{\nu}}{\partial{z}}^2\right)dr+2\rho\nu_1\nu_2dz.
\end{equation}

 Once solved, particular solutions for $\nu$ and $\lambda$ are given by
\begin{equation}
    e^{-\nu}=\left(\frac{\rho_0}{\rho}\right)^h, \quad e^{\lambda-\nu}=\left(\frac{\rho}{\rho_0}\right)^{h^2-h},
\end{equation}
\noindent where $h$ is an arbitrary real constant. Note that, with the form of line element given by \eqref{1919 LC metric}, it is possible to choose coordinates, such that $\rho_0$ may be eliminated from the metric, although this will result in $\phi$ not being parameterized from $(0,2\pi)$. This will correspond to a global angular defect discussed in \cite{krisch2011levi,maccallum2011editorial}. Since the metric is independent of the angular coordinate, such a reparameterization will not necessarily affect the local classification given by either CM or CK invariants. Explicitly, we will take this metric to be
\begin{equation}\label{19met}
    ds^2=-\rho^{2h}dt^2+\rho^{2\left(h^2-h\right)}\left(d\rho^2+dz^2\right)+\rho^{2(1-h)}dx_3^2,
\end{equation}
and use it for all calculations going forward. We note, once again for clarity, that $x_3\in[0,a]$ and $a$ is determined by the specific value of $r_0$ taken above.

Going forward we will take a null frame given by:
\begin{equation}
\begin{gathered}
    l=\frac{1}{\sqrt{2}}\left(-\rho^h\frac{\partial}{\partial t}+\rho^{h^2-h}\frac{\partial}{\partial \rho}\right), \quad
    k=\frac{1}{\sqrt{2}}\left(-\rho^h\frac{\partial}{\partial t}-\rho^{h^2-h}\frac{\partial}{\partial \rho}\right), \\
    m=\frac{1}{\sqrt{2}}\left(\rho^{h^2-h}\frac{\partial}{\partial z}+i\rho^{h-1}\frac{\partial}{\partial x_3}\right), \quad 
    \overbar{m}=\frac{1}{\sqrt{2}}\left(\rho^{h^2-h}\frac{\partial}{\partial z}-i\rho^{h-1}\frac{\partial}{\partial x_3}\right).
\end{gathered}
\end{equation}

\subsection{Curvature Invariants and CK Classification}
In the above frame, the only non-vanishing components of curvature are
\begin{equation}
    \Psi_0=\Psi_4=\frac{1}{2}h\left(h^2-1\right)\rho^{-2\left(h^2-h+1\right)}~~~\text{and}~~~ \Psi_2=-\frac{1}{2}h\left(h-1\right)^2\rho^{-2\left(h^2-h+1\right)}~.
\end{equation}

This frame is invariantly defined and thus these are the zeroth order CK invariants. Additionally, this frame fixes out all isotropy. Taking the covariant derivatives, the only non-vanishing terms are
\begin{equation}
\begin{aligned}
    &C_{kmkm;k}=-C_{\overbar{m}l\overbar{m}l;l}=\frac{\left(h^5-h\right)\rho^{-3\left(1+h^2-h\right)}}{\sqrt{2}},\\
    &C_{kmkm;l}=C_{kmkl;m}=-C_{kl\overbar{m}l;\overbar{m}}=-C_{\overbar{m}l\overbar{m}l;\overbar{m}}=-\frac{\left(h^2+h\right)\left(h-1\right)^3\rho^{-3\left(1+h^2-h\right)}}{\sqrt{2}},\\
    &C_{kmkl;\overbar{m}}=C_{kmkl;k}=-C_{km\overbar{m}l;l}=-C_{kl\overbar{m}l;m}=-\frac{\left(h-1\right)^2\left(h^3-h^2+h\right)\rho^{-3\left(1+h^2-h\right)}}{\sqrt{2}},
\end{aligned}
\end{equation}
none of which are functionally independent from the zeroth order invariants. Thus, the algorithm terminates here.

We do note that for certain special cases this spacetime is of type D $\left(h=-1,\frac{1}{2},2\right)$~\cite{stephani2009exact,herrera1999levi}. In these cases, the algorithm must be approached separately as the frame given is only invariantly defined when $h=-1$. Additionally, since the Weyl tensor is now more algebraically special, there will be boost and spin isotropy remaining at zeroth order. This means that, in general, one will need to compute higher order derivatives to fully classify these special cases. Here, we will work each of these three cases explicitly. 

When $h=-1$, the frame given by \eqref{19met} becomes
\begin{equation}
\begin{gathered}
    l=\frac{1}{\sqrt{2}}\left(-\rho^{-1}\frac{\partial}{\partial t}+\rho^{2}\frac{\partial}{\partial \rho}\right), \quad
    k=\frac{1}{\sqrt{2}}\left(-\rho^{-1}\frac{\partial}{\partial t}-\rho^{2}\frac{\partial}{\partial \rho}\right), \\
    m=\frac{1}{\sqrt{2}}\left(\rho^{2}\frac{\partial}{\partial z}+i\rho^{2}\frac{\partial}{\partial x_3}\right), \quad 
    \overbar{m}=\frac{1}{\sqrt{2}}\left(\rho^{2}\frac{\partial}{\partial z}-i\rho^{2}\frac{\partial}{\partial x_3}\right).
\end{gathered}
\end{equation}

 In this frame, the only non-vanishing curvature component is 
\begin{equation}
    \Psi_2=2\rho^{-6},
\end{equation}
where we have one functionally independent term, and remaining isotropy
\begin{equation}
    \begin{pmatrix}
    \alpha & 0 \\
    0 & \alpha^{-1}
    \end{pmatrix}
    \text{, with $\alpha\in\mathbb{C}$ and }
    \begin{pmatrix}
    0 & 1 \\
    -1 & 0
    \end{pmatrix}.
     \label{Disotropy}
\end{equation}

At first order, the non-vanishing derivatives are
\begin{equation}
    C_{kmkl;\overbar{m}}=C_{km\overbar{m}l;k}=-C_{km\overbar{m}l;l}=-C_{kl\overbar{m}l;m}=6\sqrt{2}\rho^{-9},
\end{equation}
\noindent which are not functionally independent of the zeroth order components. At first order, the isotropy group is reduced to just
\begin{equation}
    \begin{pmatrix}
    e^{i\theta} & 0 \\
    0 & e^{-i\theta}
    \end{pmatrix}
    \text{, with $\theta\in\mathbb{R}$}.
    \label{rotsym}
\end{equation}

 The non-vanishing second derivative components are
\begin{equation}
    \begin{gathered}
    \begin{aligned}
           C_{kmkm;\overbar{m}\overbar{m}}&=C_{kmkl;k\overbar{m}}=C_{kmkl;\overbar{m}k}=C_{km\overbar{m}l;kk}=C_{km\overbar{m}l;ll}=C_{kl\overbar{m}l;lm}=C_{kl\overbar{m}l;ml}\\
           &=C_{\overbar{m}l\overbar{m}l;mm}=-C_{kl\overbar{m}l;\overbar{m}m}=-C_{kmkl;l\overbar{m}}=-C_{km\overbar{m}l;\overbar{m}m}=-C_{kl\overbar{m}l;km}\\
            &=-\frac{4}{5}C_{kmkl;\overbar{m}l}=-\frac{4}{5}C_{km\overbar{m}l;kl}=-\frac{4}{5}C_{km\overbar{m}l;lk}=-\frac{4}{5}C_{kl\overbar{m}l;mk}=\frac{48}{\rho^{12}},
    \end{aligned}\\
    \end{gathered}
\end{equation}
\noindent which do not reduce the above isotropy. Thus, the algorithm stops at second order. 

In the following two cases, the functional independence follows exactly as above. At zeroth order the isotropy is identical, but in both cases, at first order, it is reduced to
\begin{equation}
    \begin{pmatrix}
    \beta & 0 \\
    0 & \beta^{-1}
    \end{pmatrix}
    \text{, with $\beta\in\mathbb{R}$}.
    \label{betaisom}
\end{equation}

For $h=\frac{1}{2}$, using the frame:
\begin{equation}
\begin{gathered}
       l=\frac{1}{\sqrt{2}}\left(-\sqrt{\rho}\frac{\partial}{\partial t}+\sqrt{\rho}\frac{\partial}{\partial x_3}\right), \quad k=\frac{1}{\sqrt{2}}\left(-\sqrt{\rho}\frac{\partial}{\partial t}-\sqrt{\rho}\frac{\partial}{\partial x_3}\right),\\
       m=\frac{1}{\sqrt{2}}\left(\rho^{-1/4}\frac{\partial}{\partial \rho}+i\rho^{-1/4}\frac{\partial}{\partial z}\right), \quad \overbar{m}=\frac{1}{\sqrt{2}}\left(\rho^{-1/4}\frac{\partial}{\partial \rho}-i\rho^{-1/4}\frac{\partial}{\partial z}\right),
\end{gathered}
\end{equation}
we have the CK invariants:
\begin{equation}
\begin{gathered}
       \Psi_2=\frac{1}{8}\rho^{-3/2},\\
       C_{kmkl;l}=C_{km\overbar{m}l;m}=C_{km\overbar{m}l;\overbar{m}}=C_{kl\overbar{m}l;k}=-\frac{3}{16\sqrt{2}}\rho^{-9/4},\\
       C_{kmkm;ll}=C_{kmkl;lm}=C_{kmkl;ml}=C_{kmkl;\overbar{m}l}=C_{km\overbar{m}l;kl}=C_{km\overbar{m}l;lk}\\
       =C_{km\overbar{m}l;mm}=C_{km\overbar{m}l;\overbar{m}\overbar{m}}=C_{kl\overbar{m}l;k\overbar{m}}=C_{kl\overbar{m}l;mk}=C_{kl\overbar{m}l;\overbar{m}k}=C_{\overbar{m}l\overbar{m}l;kk}\\
       =\frac{4}{5}C_{kmkl;l\overbar{m}}=\frac{4}{5}C_{km\overbar{m}l;m\overbar{m}}=\frac{4}{5}C_{km\overbar{m}l\overbar{m}m}=\frac{4}{5}C_{kl\overbar{m}l;km}=\frac{3}{16}\rho^{-3}.
\end{gathered}
\end{equation}

For $h=2$, using the frame:
\begin{equation}
\begin{gathered}
       l=\frac{1}{\sqrt{2}}\left(-\rho^2\frac{\partial}{\partial t}+\rho^2\frac{\partial}{\partial z}\right), \quad k=\frac{1}{\sqrt{2}}\left(-\rho^2\frac{\partial}{\partial t}-\rho^2\frac{\partial}{\partial z}\right),\\
       m=\frac{1}{\sqrt{2}}\left(\rho^2\frac{\partial}{\partial \rho}+\frac{i}{\rho}\frac{\partial}{\partial x_3}\right), \quad \overbar{m}=\frac{1}{\sqrt{2}}\left(\rho^2\frac{\partial}{\partial \rho}-\frac{i}{\rho}\frac{\partial}{\partial x_3}\right),
\end{gathered}
\end{equation}
we have the CK invariants:
\begin{equation}
\begin{gathered}
       \Psi_2=2\rho^{-6},\\
       C_{kmkl;l}=C_{km\overbar{m}l;m}=C_{km\overbar{m}l;\overbar{m}}=C_{kl\overbar{m}l;k}=-6\sqrt{2}\rho^{-9},\\
       C_{kmkm;ll}=C_{kmkl;lm}=C_{kmkl;ml}=C_{kmkl;\overbar{m}l}=C_{km\overbar{m}l;kl}=C_{km\overbar{m}l;lk}\\
       =C_{km\overbar{m}l;mm}=C_{km\overbar{m}l;\overbar{m}\overbar{m}}=C_{kl\overbar{m}l;k\overbar{m}}=C_{kl\overbar{m}l;mk}=C_{kl\overbar{m}l;\overbar{m}k}=C_{\overbar{m}l\overbar{m}l;kk}\\
       =\frac{4}{5}C_{kmkl;l\overbar{m}}=\frac{4}{5}C_{km\overbar{m}l;mm}=\frac{4}{5}C_{km\overbar{m}l;\overbar{m}m}=\frac{4}{5}C_{kl\overbar{m}l;kl}=48\rho^{-12}.
\end{gathered}
\end{equation}

 Here, we can see that the $h=1/2$ and $h=2$ solutions are in fact identical as their CK invariants can be found to be compatible. 

In the case where $h=0,1$, the spacetime is (locally) flat, and, thus, all invariants will vanish, although globally there are topological properties mentioned above not captured by this approach.

The only two non-vanishing CM invariants are
\begin{equation}
    w_1=2\left(h^2-h\right)^2\left(h^2-h+1\right)\rho^{-4\left(h^2-h+1\right)}, \quad w_2=-3\left(h^2-h\right)^4\rho^{-6\left(h^2-h+1\right)},
\end{equation}
\noindent which reduces as expected for the special values of $h$.
\subsection{Kasner Generalization}

In \cite{delice2004kasner}, a generalization of the 1919 solution was given as
\begin{equation}
    ds^2=-r^{2D}dt^2+t^{2A}dr^2+r^{2E}t^{2B}dz^2+\alpha^2r^{2F}t^{2C}d\phi^2,
\end{equation}
\noindent which is in general stationary (rather than static). The constants in this solution are related via the Kasner constraints
\begin{equation}
    A+B+C=A^2+B^2+C^2=1,\quad D+E+F=D^2+E^2+F^2=1,
\end{equation}
and take specific values given by:
\begin{equation}
\begin{gathered}
    A=\frac{2s+H}{S+H}, \quad B=\frac{\left(2s-1\right)\left(2s+\epsilon\left(2s-1\right)\right)}{S+H}, \quad C=\frac{\left(1-2s\right)\left(1+\epsilon\left(2s-1\right)\right)}{S+H},\\
    D=\frac{2s}{S}, \quad E=\frac{2s\left(2s-1\right)}{S}, \quad F=\frac{1-2s}{S},
\end{gathered}
\end{equation}
\noindent where $s$ is a real constant, $\epsilon=\pm1$, and 
\begin{equation}
    H=\epsilon\left(4s^2-1\right)+\left(1-2s\right)^2, \quad S=4s^2-2s+1.
\end{equation}

 Computing the curvature with respect to the null tetrad
\begin{equation}
\begin{gathered}
    l=\frac{1}{\sqrt{2}}\left(-r^D\frac{\partial}{\partial t}+t^A\frac{\partial}{\partial r}\right), \quad k=\frac{1}{\sqrt{2}}\left(-r^D\frac{\partial}{\partial t}-t^A\frac{\partial}{\partial r}\right),\\
    m=\frac{1}{\sqrt{2}}\left(r^Et^B\frac{\partial}{\partial z}+i\alpha^2r^Ft^C \frac{\partial}{\partial \phi}\right), \quad \overbar{m}=\frac{1}{\sqrt{2}}\left(r^Et^B\frac{\partial}{\partial z}-i\alpha^2r^Ft^C \frac{\partial}{\partial \phi}\right),
\end{gathered}    
\end{equation}
we have the non-vanishing Ricci components:
\begin{equation}
\begin{gathered}
    \Phi_{00}=-\frac{1}{2}\left(B\left(D-E\right)+C\left(D-F\right)+A\left(E+F\right)\right)t^{-2+E+F}r^{-2+B+C},\\
    \Phi_{22}=\frac{1}{2}\left(B+C+E-3BE-2CE+F-2BF-3CF\right)r^{-2+E+F}t^{-2+B+C},
\end{gathered}    
\end{equation}
and
\begin{equation}
\begin{split}
\Psi_0&=\frac{1}{2r^2t^2}\left(C\left(-2+B+2C\right)r^{2\left(E+F\right)}\right.\\
&\left.+\left(C-E+B\left(3E-1\right)+F-3CF\right)r^{E+F}t^{B+C}+F\left(-2+E+2F\right)t^{2\left(B+C\right)}\right),\\
\Psi_2&=\frac{1}{2}\left(-\frac{BCr^{-2D}}{t^2}+\frac{EFt^{-2A}}{r^2}\right),\\
\Psi_4&=\frac{1}{2r^2t^2}\left(C\left(-2+B+2C\right)r^{2\left(E+F\right)}\right.\\
&\left.+\left(B+E-3BE-F+C\left(3F-1\right)\right)r^{E+F}t^{B+C}+F\left(-2+E+2F\right)t^{2\left(B+C\right)}\right).
\end{split}
\end{equation}

 This frame is not canonical, as this spacetime is Petrov type I, and $\Psi_0\neq\Psi_4$, but the boost 
\begin{equation}
       \begin{pmatrix}
    \left(\Psi_0 /\Psi_4\right)^{1/8} & 0 \\
    0 & \left(\Psi_4 /\Psi_0\right)^{1/8}
    \end{pmatrix},
\end{equation}
\noindent will bring it to the canonical form
\begin{equation}
\begin{gathered}
    \Psi_0'=\Psi_4'=\sqrt{\Psi_0\Psi_4},\\
    \Psi_2'=\Psi_2.
\end{gathered}
\end{equation}

 For compactness we will write curvature components of the new canonical frame in terms of the old frame, where the new frame will be primed and the old frame unprimed. 
At zeroth order we have no remaining isotropy and have only two functionally independent terms. The first order derivatives, shown below, cannot reduce the isotropy further and contain no new functionally independent terms, and so the algorithm terminates. 

\vspace{-10pt}
\begin{gather*}
\begin{gathered}
    C_{kmkm;k}=\frac{1}{2\sqrt{2}r^3t^3}\left(2C\left(4-3B-5C+2BC+C^2\right)r^{3(E+F)}+(4(E-F)\right.\\
    +B\left(1-6E+C\left(7-4E-13F\right)+3F\right)+C\left(-15+14E+C\left(14-17E-17F\right)\right.\\
    \left.+23F)\right)r^{2(E+F)}t^{B+C}+\left(-E+15F-7F\left(E+2F\right)+C\left(4+E\left(-3+13F\right)\right.\right.\\
    \left.+F\left(-23+17F\right))+B\left(-4+E\left(6+4F\right)+F\left(-14+17F\right)\right)\right)r^{E+F}t^{2(B+C)}\\
    \left.-2F\left(4-3E-5F+2EF+F^2\right)t^{3(B+C)}\right)\left(\frac{\Psi_{4}}{\Psi_{0}}\right)^{3/4},
\end{gathered}\\
\begin{gathered}
    C_{kmkm;l}=-\frac{1}{2\sqrt{2}r^3t^3}\left(2C^{ }\left(\left(-1+C\right)C+B\left(-1+2C\right)\right)r^{3(E+F)}+\left(C\left(-1+2E\right.\right.\right.\\
    \left.\left.+C\left(2-3E-3F\right)+F\right)+B\left(-1+C+2E+F-3CF\right)\right)r^{2(E+F)}t^{B+C}\\
    +\left(E\left(-1+2B+C+F-3CF\right)+F\left(-1+2B+C+2F-3\left(B+C\right)F\right)\right)r^{E+F}t^{2(B+C)}\\
    \left.+2F\left(\left(-1+F\right)F+E\left(-1+2F\right)\right)t^{3(B+C)}\right)\left(\frac{\Psi_{4}}{\Psi_{0}}\right)^{1/4},\\\\
\end{gathered}\\
\begin{gathered}
    C_{kmkl;m}=\frac{1}{4\sqrt{2}r^3t^3}\left(-4C\left(\left(-1+C\right)C+B\left(-1+2C\right)\right)r^{3(E+F)}\right.\\
\end{gathered}\\
\begin{gathered}
    +\left(-B\left(-1+C+2E+4CE+F-5CF\right)
    +C\left(1+F+C\left(-2+E+F\right)\right)\right)\\
    \times r^{2(E+F)}t^{B+C}+\left(F\left(1+C+\left(-2+B+C\right)F\right)-E\left(-1+C+F-5CF\right.\right.\\
    \left.\left.\left.+B\left(2+4F\right)\right)\right)r^{E+F}t^{2(B+C)}-4F\left(\left(-1+F\right)F+E\left(-1+2F\right)\right)t^{3(B+C)}\right)\left(\frac{\Psi_{4}}{\Psi_{0}}\right)^{1/4},\\\\
\end{gathered}\\
\begin{gathered}
    C_{kmkl;\overbar{m}}=\frac{1}{4\sqrt{2}}r^{-3+E+F}t^{-3+B+C}((B+C-2BE+(-2+C)CE\\
    -BF-C(1+C)F+BC(-3+2E+F))r^{E+F}+4BCr^{2(E+F)}t^{-B-C}\\
    +\left(E-CE+2BE\left(-1+F\right)+F+\left(-3+C\right)EF+B\left(-2+F\right)F-CF\left(1+F\right)\right)\\
    \left.\times t^{B+C} +4EFr^{-E-F}t^{2(B+C)}\right)\left(\frac{\Psi_{4}}{\Psi_{0}}\right)^{1/4},\\\\
\end{gathered}\\
\begin{gathered}
    C_{km\overbar{m}l;k}=\frac{1}{\sqrt{2}}r^{-3(1+D)}t^{-3(1+A)}\left(EFr^{3D}t^3+BCr^3t^{3A}\right.\\
    \left.+\left(-1+B+C\right)EFr^{1+2D}t^{2+A}+BC\left(-1+E+F\right)r^{2+D}t^{1+2A}\right)\left(\frac{\Psi_{4}}{\Psi_{0}}\right)^{1/4},\\\\
    C_{km\overbar{m}l;l}=\frac{r^{-3(1+D)}t^{-3(1+A)}}{\sqrt{2}\left(\frac{\Psi_{4}}{\Psi_{0}}\right)^{1/4}}\left(-EFr^{3D}t^3\right.
    +BCr^3t^{3A}\\
    +\left(-1+B+C\right)EFr^{1+2D}t^{2+A}-BC\left(-1+E+F\right)\left.r^{2+D}t^{1+2A}\right),\\\\
\end{gathered}\\
\begin{gathered}
    C_{kl\overbar{m}l;m}=\frac{r^{-3+E+F}t^{-3+B+C}}{4\sqrt{2}\left(\frac{\Psi_{4}}{\Psi_{0}}\right)^{1/4}}\left(-\left(B+C-2BE+\left(-2+C\right)CE-BF-C\left(1+C\right)F\right.\right.\\
    \left.+BC\left(-3+2E+F\right)\right)r^{E+F}+4BCr^{2(E+F)}t^{-B-C}+\left(E-CE+2BE\left(-1+F\right)\right.\\
    \left.\left.+F+\left(-3+C\right)EF+B\left(-2+F\right)F-CF\left(1+F\right)\right)t^{B+C}-4EFr^{-E-F}t^{2(B+C)}\right),\\\\
\end{gathered}\\
\begin{gathered}
    C_{kl\overbar{m}l;\overbar{m}}=\frac{1}{4\sqrt{2}r^3t^3\left(\frac{\Psi_{4}}{\Psi_{0}}\right)^{1/4}}\left(-4C\left(\left(-1+C\right)C+B\left(-1+2C\right)\right)r^{3(E+F)}\right.\\
    +\left(B\left(-1+C+2E+4CE+F-5CF\right)-C\left(1+F+C\left(-2+E+F\right)\right)\right)r^{2(E+F)}t^{B+C}\\
    +\left(F\left(1+C+\left(-2+B+C\right)F\right)-E\left(-1+C+F-5CF+B\left(2+4F\right)\right)\right)r^{E+F}t^{2(B+C)}\\
    \left.+4F\left(\left(-1+F\right)F+E\left(-1+2F\right)\right)t^{3(B+C)}\right),\\\\
\end{gathered}\\
\begin{gathered}
    C_{\overbar{m}l\overbar{m}l;k}=\frac{1}{2\sqrt{2}r^3t^3\left(\frac{\Psi_{4}}{\Psi_{0}}\right)^{1/4}}\left(-2C\left(\left(-1+C\right)C+B\left(-1+2C\right)\right)r^{3(E+F)}\right.\\
    +\left(C\left(-1+2E+C\left(2-3E-3F\right)+F\right)+B\left(-1+C+2E+F-3CF\right)\right)r^{2(E+F)}t^{B+C}\\
    +\left(-E\left(-1+2B+C+F-3CF\right)+F\left(1-2B-C-2F+3\left(B+C\right)F\right)\right)r^{E+F}t^{2(B+C)}\\
    \left.+2F\left(\left(-1+F\right)F+E\left(-1+2F\right)\right)t^{3(B+C)}\right),\\\\
\end{gathered}\\
\begin{gathered}
    C_{\overbar{m}l\overbar{m}l;l}=\frac{1}{2\sqrt{2}r^3t^3\left(\frac{\Psi_{4}}{\Psi_{0}}\right)^{3/4}}\left(2C\left(4-3B-5C+2BC+C^2\right)r^{3(E+F)}\right.\\
    +\left(-4E+4F+B\left(-1+6E-3F+C\left(-7+4E+13F\right)\right)+C\left(15-14E-23F\right.\right.\\
    \left.\left.+C\left(-14+17E+17F\right)\right)\right)r^{2(E+F)}t^{B+C}+\left(-E+15F-7F\left(E+2F\right)\right.\\
    +C\left(4+E\left(-3+13F\right)+F\left(-23+17F\right)\right)+B\left(-4+E\left(6+4F\right)\right.\\
    \left.\left.\left.+F\left(-14+17F\right)\right)\right)r^{E+F}t^{2(B+C)}+2F\left(4-3E-5F+2EF+F^2\right)t^{3(B+C)}\right).
\end{gathered}
\stepcounter{equation}
\tag{\arabic{equation}}
\end{gather*}

Here, the non-vanishing CM invariants are:

\vspace{-10pt}
\begin{gather*}
    \begin{gathered}
    r_1=\sqrt{2r_3}=-\frac{1}{2}\left(B+C+E-3BE-2CE+F-2BF-3CF\right)\\
    \times\left(B\left(D-E\right)+C\left(D-F\right)+A\left(E+F\right)\right)r^{2\left(E+F-2\right)}t^{2\left(B+C-2\right)},
    \end{gathered}\\\\
    \begin{gathered}
    w_1=\frac{1}{2r^4t^4}\left(3\left(BCr^{2\left(E+F\right)}-EFt^{2\left(B+C\right)}\right)^2+\left(C\left(B+2C-2\right)r^{2\left(E+F\right)}\right.\right.\\
    +\left.\left(C-E+B\left(3E-1\right)+F-3CF\right)r^{E+F}t^{B+C}+F\left(E+2F-2\right)t^{2\left(B+C\right)}\right)\\
    \times\left(C\left(B+2C-2\right)r^{2\left(E+F\right)}+\left(B+E-3BE-F+C\left(3F-1\right)\right)r^{E+F}t^{B+C}+\right.\\
   \left.\left.+ F\left(E+2F-2\right)t^{2\left(B+C\right)}\right)\right),
   \end{gathered}\\\\
   \begin{gathered}
   w_2=\frac{3}{4r^6t^6}\left(\left(BCr^{2(E+F)}-EFt^{2(B+C)}\right)^3-\left(BCr^{2(E+F)}-EFt^{2(B+C)}\right)\right.\\
   \times\left(C\left(-2+B+2C\right)r^{2(E+F)}+\left(C-E+B\left(-1+3E\right)\right.\right.\\
   \left.\left.+F-3CF\right)r^{E+F}t^{B+C}+F\left(-2+E+2F\right)t^{2(B+C)}\right)\\
   \times\left(C\left(-2+B+2C\right)r^{2(E+F)}+\left(B+E-3BE-F+C\left(-1+3F\right)\right)r^{E+F}t^{B+C}\right.\\
   \left.\left.+F\left(-2+E+2F\right)t^{2(B+C)}\right)\right),
   \end{gathered}\\\\
   \begin{gathered}
    m_1=\frac{1}{4}\left(B\left(E-D\right)+C\left(F-D\right)-A\left(E+F\right)\right)\\
   \times\left(-E-F+B\left(-1+3E+2F\right)+C\left(-1+2E+3F\right)\right)r^{4\left(-2+E+F\right)}t^{4\left(-2+B+C\right)}\\
   \times\left(-EFr^{2D}t^2+BCr^2t^{2A}\right),
   \end{gathered}\\\\
   \begin{gathered}
   m_2=\frac{1}{8}\left(B+C+E-3BE-2CE+F-2BF-3CF\right)\\
   \times\left(B\left(D-E\right)+C\left(D-F\right)+A\left(E+F\right)\right)r^{2(-4+E+F)}t^{2(-4+B+C)}\\
   \times\left(-\left(BCr^{2(E+F)}-EFt^{2(B+C)}\right)^2-\left(C\left(-2+B+2C\right)r^{2(E+F)}\right.\right.\\
   \left.+\left(C-E+B\left(-1+3E\right)+F-3CF\right)r^{E+F}t^{B+C}+F\left(-2+E+2F\right)t^{2(B+C)}\right)\\
   \times\left(C\left(-2+B+2C\right)r^{2(E+F)}+\left(B+E-3BE-F+C\left(-1+3F\right)\right)r^{E+F}t^{B+C}\right.\\
   \left.\left.+F\left(-2+E+2F\right)t^{2(B+C)}\right)\right),
   \end{gathered}\\\\
   \begin{gathered}
    m_3=\frac{1}{16}r^{2\left(E+F-4\right)}t^{2\left(B+C-4\right)}\left(-2\left(B\left(E-D\right)+C\left(F-D\right)-A\left(E+F\right)\right)\right.\\
    \times\left(-E-F+B\left(-1+3E+2F\right)+C\left(-1+2E+3F\right)\right)r^{-4D}t^{-4A}\left(EFr^{2D}t^2-BCr^2t^{2A}\right)^2\\
    +\left(B+C+E-3BE-2CE+F-2BF-3CF\right)^2\\
    \times\left(2C^2r^{2\left(E+F\right)}+Cr^{E+F}\left(\left(B-2\right)r^{E+F}+\left(1-3F\right)t^{B+C}\right)\right.\\
    \left.+t^{B+C}\left(B\left(3E-1\right)r^{E+F}+F\left(r^{E+F}+2\left(F-1\right)t^{B+C}\right)+E\left(-r^{E+F}+Ft^{B+C}\right)\right)\right)^2\\
    +\left(B\left(D-E\right)+C\left(D-F\right)+A\left(E+F\right)\right)^2\\
    \times\left(2C^2r^{2\left(E+F\right)}+Cr^{E+F}\left(\left(B-2\right)r^{E+F}+\left(3F-1\right)t^{B+C}\right)\right.\\
    \left.\left.+t^{B+C}\left(-B\left(3E-1\right)r^{E+F}-Fr^{E+F}+2\left(F-1\right)Ft^{B+C}+E\left(r^{E+F}+Ft^{B+C}\right)\right)\right)^2\right),
   \end{gathered}\\\\
   \begin{gathered}
   m_5=\frac{1}{8}r^{2\left(E+F-4\right)}t^{2\left(B+C-4\right)}\left(\frac{1}{2}\left(B\left(E-D\right)+C\left(F-D\right)-A\left(E+F\right)\right)\right.\\
   \times\left(-E-F+B\left(3E+2F-1\right)+C\left(2E+3F-1\right)\right)r^{-2-6D}t^{-2-6A}\left(BCr^2t^{2A}-EFr^{2D}t^2\right)^3\\
   +\frac{1}{2}\left(B+C+E-3BE-2CE+F-2BF-3CF\right)^2r^{2\left(E+F-2\right)}t^{2\left(B+C-2\right)}\\
   \times\left(EFr^{2D}t^2-BCr^2t^{2A}\right)\left(2C^2r^{2\left(E+F\right)}+Cr^{E+F}\left(\left(B-2\right)r^{E+F}+\left(1-3F\right)t^{B+C}\right)\right.\\
   \left.+t^{B+C}\left(B\left(3E-1\right)r^{E+F}+F\left(r^{E+F}+2\left(F-1\right)t^{B+C}\right)+E\left(-r^{E+F}+Ft^{B+C}\right)\right)\right)^2\\
   +\frac{1}{2}\left(B\left(E-D\right)+C\left(F-D\right)-A\left(E+F\right)\right)\left(-E-F+B\left(3E+2F-1\right)\right.\\
   \left.+C\left(2E+3F-1\right)\right)r^{-2\left(1+D\right)}t^{-2\left(1+A\right)}\left(-EFr^{2D}t^2+BCr^2t^{2A}\right)\left(2C^2r^{2\left(E+F\right)}\right.\\
   +Cr^{E+F}\left(\left(B-2\right)r^{E+F}+\left(1-3F\right)t^{B+C}\right)+t^{B+C}\left(B\left(3E-1\right)r^{E+F}\right.\\
   \left.\left.+F\left(r^{E+F}+2\left(-1+F\right)t^{B+C}\right)+E\left(-r^{E+F}+Ft^{B+C}\right)\right)\right)\\
    \times\left(2C^2r^{2\left(E+F\right)}+Cr^{E+F}\left(\left(-2+B\right)r^{E+F}+\left(3F-1\right)t^{B+C}\right)+t^{B+C}\left(-B\left(3E-1\right)r^{E+F}\right.\right.\\
   \left.\left.-Fr^{E+F}+2\left(F-1\right)Ft^{B+C}+E\left(r^{E+F}+Ft^{B+C}\right)\right)\right)\\
   \left.-\frac{1}{2}\left(B\left(D-E\right)+C\left(D-F\right)+A\left(E+F\right)\right)^2r^{2\left(-2+E+F\right)}t^{2\left(-2+B+C\right)}\right.\\
   \times\left(-EFr^{2D}t^2+BCr^2t^{2A}\right)\left(2C^2r^{2\left(E+F\right)}+Cr^{E+F}\left(\left(-2+B\right)r^{E+F}+\left(-1+3F\right)t^{B+C}\right)\right.\\
   \left.\left.+t^{B+C}\left(-B\left(3E-1\right)r^{E+F}-Fr^{E+F}+2\left(F-1\right)Ft^{B+C}+E\left(r^{E+F}+Ft^{B+C}\right)\right)\right)^2\right),\\
   \end{gathered}
   \stepcounter{equation}
   \tag{\arabic{equation}}
\end{gather*}

\noindent where we point out that the vanishing of the Ricci invariant $r_2$ and the mixed invariant $m_4$ are useful for identifying this spacetime. 
\section{The Longitudinal, Quadrantal, and Oblique Levi-Civita Solutions (1918)}
There are three classes of solutions that are less commonly discussed in the literature than the  proceeding  two  we  presented. These solutions are derived in \cite{levi1918ds2quadrantali,levi1918ds2oblique,levi1918ds2longitudinali} and all three are reviewed and summarized in \cite{levi1918ds2oblique}. These three papers are not published in English and the latter two remain nearly uncited directly, thus we will carefully review these solutions in more depth. Throughout this section we will also present the non-vanishing $\mathcal{I}$ invariants (see \cite{abdelqader2015invariant}) as these scalars contain information not present in the CM invariants, due to being constructed from derivatives of the Weyl tensor.
\subsection{The Longitudinal Solutions}
Here we characterize the solutions described in \cite{levi1918ds2longitudinali}, which Levi-Civita refers to as longitudinal solutions, described by the metric
\begin{equation}
    ds^2=-\left(\mu-\epsilon\eta\right)dt^2+\frac{d\sigma^2}{\eta^2}+\frac{d\eta^2}{K_0\eta^4\left(\mu-\epsilon\eta\right)},
\end{equation}
\noindent where $d\sigma^2$ is the line element for a two-dimensional space with constant Gaussian curvature and $\mu$, $\eta$ are coordinates which run over intervals such that the spacetime is consistent with the chosen signature convention, $\mu$ is a real constant, $K_0$ is a positive constant, and $\epsilon=\pm1$.

This solution can be split into three distinct subcases depending on the sign of $\mu$. Here, we will move to work with the coordinates given by \cite{jordan2009republication}, as they are closer to those typically employed in modern references and explicitly reduce the number of free constants to one. The three forms given are
\begin{align}
  \left(\mu>0\right):\quad  ds^2&=-\left(1-\frac{2m}{r}\right)dt^2+\left(1-\frac{2m}{r}\right)^{-1}dr^2+r^2\left(d\theta^2+\sin(\theta)^2d\phi^2\right)\label{3222},\\
  &\text{where $m>0$ and $2m<r<\infty$, or $m<0$ and $0<r<\infty$},\notag\\ 
  \left(\mu<0\right):\quad   ds^2&=-\left(\frac{2m}{z}-1\right)dt^2+\left(\frac{2m}{z}-1\right)^{-1}dz^2+z^2\left(dr^2+\sinh(r)^2d\phi^2\right)\label{3223},\\
  &\text{where $m>0$ and $0<z<2m$},\notag\\
  \left(\mu=0\right):\quad   ds^2&=-\frac{dt^2}{z}+zdz^2+z^2\left(dr^2+r^2d\phi^2\right)\label{3224},\\
  &\text{where $z>0$}\notag.
\end{align}

 For completeness, we note that \eqref{3224} is not the only solution for $\mu=0$ (we also note that the degenerate static vacuum fields are also listed in Table 2--3.1 in \cite{witten1962gravitation}, in which equation \eqref{3222} is classified as ``A1",
\eqref{3223} is classified as ``A2", and \eqref{3224} is classified as ``A3", where $b=2m$).
 In particular, this solution takes Gaussian flat two-spaces to be a two-dimensional plane. An equally valid choice would be to take these two spaces to be cylinders,
\begin{equation}
    ds^2=-\frac{dt^2}{z}+zdz^2+z^2\left(d\rho^2+a^2d\phi^2\right),
\end{equation}
but these solutions are only different globally, and, thus, both the CK algorithm and CM invariants will not be able to detect this difference.

The solution given by \eqref{3222} is the Schwarzschild solution which has been invariantly characterized via the CK algorithm in \cite{karlhede1980review}, via CM invariants in \cite{mattingly2020curvature}, and via $\mathcal{I}$ invariants as a subcase in \cite{abdelqader2015invariant}. The solutions given by \eqref{3223} and \eqref{3224} are distinct and, as such, we will explicitly state CK and scalar invariants. 

For \eqref{3223}, the null frame:
\begin{equation}
\begin{gathered}
    l=\frac{1}{\sqrt{2}}\left(-\sqrt{\frac{2m}{z}-1}\frac{\partial}{\partial t}+\frac{1}{\sqrt{\frac{2m}{z}-1}}\frac{\partial}{\partial z}\right),\\ k=\frac{1}{\sqrt{2}}\left(-\sqrt{\frac{2m}{z}-1}\frac{\partial}{\partial t}-\frac{1}{\sqrt{\frac{2m}{z}-1}}\frac{\partial}{\partial z}\right),\\
    m=\frac{1}{\sqrt{2}}\left(z\frac{\partial}{\partial r}+iz\sinh(r)\frac{\partial}{\partial \phi}\right), \quad \overbar{m}=\frac{1}{\sqrt{2}}\left(z\frac{\partial}{\partial r}-iz\sinh(r)\frac{\partial}{\partial \phi}\right),
\end{gathered}
\end{equation}
\noindent gives the only non-vanishing curvature component to be
\begin{equation}
    \Psi_2=mz^{-3}.
\end{equation}

This frame is, therefore, invariant to zeroth order, and the remaining isotropy is given by \eqref{Disotropy}. The independent, first order, non-zero derivatives (components) are
\begin{equation}
    C_{kmkl;\overbar{m}}=C_{km\overbar{m}l;k}=-C_{km\overbar{m}l;l}=-C_{kl\overbar{m}l;m}=\frac{3m}{\sqrt{2}z^4}\sqrt{\frac{2m}{z}-1}, \label{longderiv1}
\end{equation}
\noindent which are not functionally independent of the zeroth order components. At first order, the isotropy group is reduced to just \eqref{rotsym}. Since the isotropy group has been reduced at first order, the algorithm proceeds to second order, where the independent non-vanishing terms are
\vspace{-6pt}
\begin{equation}
\begin{aligned}
&\begin{aligned}
    C_{kmkl;l\overbar{m}}&=C_{km\overbar{m}l;m\overbar{m}}=C_{km\overbar{m}l;m\overbar{m}}=C_{kl\overbar{m}l;km}=-C_{kmkm;\overbar{m}\overbar{m}}=-C_{kmkl;k\overbar{m}}=-C_{kmkl;\overbar{m}k}\\
    &=-C_{km\overbar{m}l;kk}=-C_{km\overbar{m}l;ll}=-C_{kl\overbar{m}l;lm}=-C_{kl\overbar{m}l;ml}=-C_{\overbar{m}l\overbar{m}l;mm}=\frac{3m\left(2z-4m\right)}{z^6},
\end{aligned}\\
    &C_{kmkl;\overbar{m}l}=C_{km\overbar{m}l;kl}=C_{km\overbar{m}l;lk}=C_{kl\overbar{m}l;mk}=\frac{3m\left(2z-5m\right)}{z^6},
\end{aligned}
\end{equation}
which possess the same isotropy and produces no new functionally independent terms. Thus the zeroth, first, and second order terms are the CK invariants needed to fully characterize this spacetime. 

In this case, the non-vanishing CM invariants are:
\begin{equation}
    w_1=-\frac{1}{6\sqrt w_2}=-\frac{z^3}{m},
\end{equation}
\noindent and the non-vanishing $\mathcal{I}$ invariants are
\begin{equation}
\begin{gathered}
    I_1=48\frac{m^2}{z^6}, \quad I_3=720\frac{m^2\left(2m-z\right)}{z^9}, \quad I_5=82944\frac{m^4\left(2m-z\right)}{z^{15}}.
\end{gathered}
\end{equation}

 We note in this case that the $I_3$ and $I_5$ invariants both vanish for the extremal value $z=2m$, which indicates this hypersurface is invariantly defined in much the same way that the Schwarzschild solution's event horizon is. 

For the spacetime given by \eqref{3224}, the algorithm proceeds identically to the preceding case with the null frame given by
\begin{equation}
\begin{gathered}
    l=\frac{1}{\sqrt{2}}\left(-\frac{1}{\sqrt{z}}\frac{\partial}{\partial t}+\sqrt{z}\frac{\partial}{\partial z}\right), \quad k= \frac{1}{\sqrt{2}}\left(-\frac{1}{\sqrt{z}}\frac{\partial}{\partial t}-\sqrt{z}\frac{\partial}{\partial z}\right),\\
    m= \frac{1}{\sqrt{2}}\left(z\frac{\partial}{\partial r}+izr\frac{\partial}{\partial \phi}\right), \quad 
    \overbar{m}= \frac{1}{\sqrt{2}}\left(z\frac{\partial}{\partial r}-izr\frac{\partial}{\partial \phi}\right),
\end{gathered}
\end{equation}
which gives CK invariants at zeroth, first, and second order as:
\begin{equation}
\begin{gathered}
    \Psi_2=\frac{1}{2}z^{-3},\\
    C_{kmkl;\overbar{m}}=C_{km\overbar{m}l;k}=-C_{km\overbar{m}l;l}=-C_{kl\overbar{m}l;m}=\frac{3}{2\sqrt{2}}z^{-9/2},\\
    \begin{aligned}
   C_{kmkm;\overbar{m}\overbar{m}}&=C_{kmkl;k\overbar{m}}=C_{kmkl;\overbar{m}k}=C_{km\overbar{m}l;kk}=C_{km\overbar{m}l;ll}=C_{kl\overbar{m}l;lm}=C_{kl\overbar{m}l;ml}\\
    &=C_{\overbar{m}l\overbar{m}l;mm}=-C_{kmkl;l\overbar{m}}=-C_{km\overbar{m}l;m\overbar{m}}=-C_{km\overbar{m}l;m\overbar{m}}=-C_{kl\overbar{m}l;km}\\
    &=-\frac{4}{5}C_{kmkl;\overbar{m}l}=-\frac{4}{5}C_{km\overbar{m}l;kl}=-\frac{4}{5}C_{km\overbar{m}l;lk}=-\frac{4}{5}C_{kl\overbar{m}l;mk}=\frac{3}{z^6}.\\
    \end{aligned}\\
\end{gathered}
\end{equation}

 A particularly interesting remark is that this special case coincides (at least locally) with \eqref{1919 LC metric} where $h=-1$. 

We also have the following non-vanishing CM invariants:
\begin{equation}
    w_1=\frac{3}{2z^6}, \quad w_2=-\frac{3}{4z^9},
\end{equation}
\noindent and $\mathcal{I}$ invariants
\begin{equation}
    I_1=\frac{12}{z^6}, \quad I_3=\frac{180}{z^9}, \quad I_5=\frac{5184}{z^{15}}.
\end{equation}
\subsection{Quadrantal Solutions}
In \cite{levi1918ds2quadrantali}, Levi-Civita gave a second set of vacuum solutions. These are similar to, but distinct from, the longitudinal solutions (which he called quadrantal solutions). He gave the following line element  
\begin{equation}
    ds^2=-\frac{e^{2\zeta\left(\psi\right)}}{\xi^2}dt^2+\frac{1}{K_0\xi^2}\left(\frac{d\xi^2}{\Xi(\xi)}+\Xi(\xi) d\phi^2+d\psi^2\right),
\end{equation}
\noindent where $\Xi\left(\xi\right)=\mu\xi^3+\epsilon\xi^2$, $K_0$ is a positive constant, $\mu$ is a real constant, and $\epsilon=\pm1$. Note that the coordinate range of $\xi$ is restricted to the subset of $\mathbb{R^+}$ such that $\Xi(\xi)>0$. The undetermined function $\zeta\left(\psi\right)$ is defined by the differential equation
\begin{equation}
    \frac{\partial^2}{\partial\psi^2}e^{\zeta\left(\psi\right)}+\mu e^{\zeta\left(\psi\right)}=0,
\end{equation}
\noindent which will have solutions of the form
\begin{equation}
\begin{split}
    \mu>0: \quad e^{\zeta\left(\psi\right)}&=\cos(\sqrt{\mu}\psi), \\
    \mu<0: \quad e^{\zeta\left(\psi\right)}&=\cosh(\sqrt{\lvert\mu\rvert}\psi), \\
    \mu=0: \quad e^{\zeta\left(\psi\right)}&=\psi.
\end{split}    
\end{equation}

 Working in the null frame
 
\begin{equation}
\begin{gathered}
    l=\frac{1}{\sqrt{2}}\left(-\frac{e^{\zeta\left(\psi\right)}}{\xi}\frac{\partial}{\partial t}+\frac{1}{\sqrt{K_0}\xi}\frac{\partial}{\partial \psi}\right), \quad k=\frac{1}{\sqrt{2}}\left(-\frac{e^{\zeta\left(\psi\right)}}{\xi}\frac{\partial}{\partial t}-\frac{1}{\sqrt{K_0}\xi}\frac{\partial}{\partial \psi}\right),\\
    m=\frac{1}{\sqrt{2}}\left(\frac{1}{\xi\sqrt{K_0\Xi\left(\xi\right)}}\frac{\partial}{\partial \xi}+\frac{i}{\xi}\sqrt{\frac{\Xi\left(\xi\right)}{K_0}}\frac{\partial}{\partial \psi}\right), \quad \overbar{m}=\frac{1}{\sqrt{2}}\left(\frac{1}{\xi\sqrt{K_0\Xi\left(\xi\right)}}\frac{\partial}{\partial \xi}-\frac{i}{\xi}\sqrt{\frac{\Xi\left(\xi\right)}{K_0}}\frac{\partial}{\partial \psi}\right),
\end{gathered}
\end{equation}
\noindent the only non-vanishing curvature component is 
\begin{equation}
    \Psi_2=\frac{1}{2}\epsilon K_0 \xi^3,
\end{equation}
\noindent which is in an invariant form to first order. The remaining isotropy is given by \eqref{Disotropy}. The independent first derivative components are
\begin{equation}
    C_{kmkl;l}=C_{km\overbar{m}l;m}=C_{km\overbar{m}l;\overbar{m}}=C_{kl\overbar{m}l;k}=\frac{3\epsilon K_0^{\frac{3}{2}}\xi^4}{2\sqrt{2}}\sqrt{\mu+\epsilon\xi}, \label{quadderiv1}
\end{equation}
\noindent which are not functionally independent of $\Psi_2$ and the isotropy is reduced to \eqref{betaisom}. The second derivative components are
\begin{equation}
\begin{aligned}
&\begin{aligned}
C_{kmkm;ll}&=C_{kmkl;lm}=C_{kmkl;ml}=C_{kmkl;\overbar{m}l}=C_{km\overbar{m}l;kl}=C_{km\overbar{m}l;lk}=C_{km\overbar{m}l;mm}\\
&=C_{km\overbar{m}l;\overbar{m}\overbar{m}}=C_{kl\overbar{m}l;k\overbar{m}}=C_{kl\overbar{m}l;mk}=C_{kl\overbar{m}l;\overbar{m}k}=C_{\overbar{m}l\overbar{m}l;kk}=3\epsilon K_0^2\xi^5\left(\mu+\epsilon\xi\right),\\
\end{aligned}\\
&\begin{aligned}
C_{kmkl;l\overbar{m}}=C_{km\overbar{m}l;m\overbar{m}}=C_{km\overbar{m}l;m\overbar{m}}=C_{kl\overbar{m}l;km}=\frac{3}{4}\epsilon K_0^2 \xi^5\left(4\mu+5\epsilon\xi\right),
\end{aligned}\\
\end{aligned}
\end{equation}
which have the same isotropy and introduce no new functionally independent terms. Thus the algorithm terminates here. In the special case where $\mu=0$ and $\epsilon=1$, this solution will coincide with \eqref{1919 LC metric} where $h=1/2$ or $h=2$.

In this case, the non-vanishing CM invariants are given by
\begin{equation}
    w_1=\frac{3}{2}K_0^2\xi^6, \quad w_2=-\frac{3}{4}\epsilon K_0^3 \xi^9,
\end{equation}
\noindent and the non-vanishing $\mathcal{I}$ invariants are given by
\begin{equation}
    I_1=12 K_0^2 \xi^6, \quad I_3=180K_0^3\xi^8\left(\mu+\epsilon\xi\right), \quad I_5=5184 K_0^5\xi^{14}\left(\mu+\epsilon\xi\right).
\end{equation}

 Once again the derivative $\mathcal{I}$ invariants will vanish at the extremal value of the coordinate $\xi$ (if that coordinate is restricted due to the signs of $\mu$ and $\epsilon$).  

We note that, despite this solution appearing very similar to the longitudinal solutions, the fact that the first order isotropy groups are distinct guarantees that these are inequivalent solutions.

\subsection{Oblique Solutions}
In \cite{levi1918ds2oblique}, a third vacuum solution was given which Levi-Civita described as oblique. This solution is characterized by the line element:
\begin{equation}
    ds^2=-\frac{H(\eta)}{\left(\xi+\eta\right)^2}dt^2+\frac{1}{K_0\xi^2}\left(\frac{d\xi^2}{\Xi(\xi)}+\frac{d\eta^2}{H(\eta)}+\Xi(\xi) d\phi^2\right),
\end{equation}
\noindent where $K_0$ is a positive constant and $H$ and $\Xi$ are given by 
\begin{equation}
    \Xi(\xi)=4\xi^3-g_2\xi-g_3, \quad H(\eta)=4\eta^3-g_2\eta+g_3,
\end{equation}
\noindent with $g_2$ and $g_3$ being real numbers. Working in the tetrad
\begin{equation}
\begin{gathered}
    l=\frac{1}{\sqrt{2}}\left(-\frac{\sqrt{g_3-g_2\eta+4\eta^3}}{\eta+\xi}\frac{\partial}{\partial t}+\frac{1}{\left(\eta+\xi\right)\sqrt{K_0\left(g_3-g_2\eta+4\eta^3\right)}}\frac{\partial}{\partial \eta}\right),\\
    k=\frac{1}{\sqrt{2}}\left(-\frac{\sqrt{g_3-g_2\eta+4\eta^3}}{\eta+\xi}\frac{\partial}{\partial t}-\frac{1}{\left(\eta+\xi\right)\sqrt{K_0\left(g_3-g_2\eta+4\eta^3\right)}}\frac{\partial}{\partial \eta}\right),\\
    m=\frac{1}{\sqrt{2K_0}}\left(\frac{1}{\left(\eta+\xi\right)\sqrt{\left(-g_3-g_2\xi+4\xi^3\right)}}\frac{\partial}{\partial \xi}+i\frac{\sqrt{-g_3-g_2\xi+4\xi^3}}{\left(\eta+\xi\right)}\frac{\partial}{\partial \phi}\right),\\
    \overbar{m}=\frac{1}{\sqrt{2K_0}}\left(\frac{1}{\left(\eta+\xi\right)\sqrt{\left(-g_3-g_2\xi+4\xi^3\right)}}\frac{\partial}{\partial \xi}-i\frac{\sqrt{-g_3-g_2\xi+4\xi^3}}{\left(\eta+\xi\right)}\frac{\partial}{\partial \phi}\right),
\end{gathered}
\end{equation}
\noindent the only non-vanishing curvature component is
\begin{equation}
    \Psi_2=2K_0\left(\eta+\xi\right)^3.
\end{equation}

 This frame is canonical and has remaining isotropy given by \eqref{Disotropy} as before. The first order derivative components are
\begin{equation}
\begin{gathered}
   C_{kmkl;l}=C_{km\overbar{m}l;m}=C_{km\overbar{m}l;\overbar{m}}=C_{kl\overbar{m}l;k}=3\sqrt{2}\sqrt{K_0^3\left(\eta+\xi\right)^6\left(-g_3-g_2\xi+4\xi^3\right)},\\
   C_{km\overbar{m}l;l}=C_{kl\overbar{m}l;m}=-C_{kmkl;\overbar{m}}=-C_{km\overbar{m}l;k}=3\sqrt{2}\sqrt{K_0^3\left(\eta+\xi\right)^6\left(g_3-g_2\eta+4\eta^3\right)}.
\end{gathered}    
\end{equation}

 These terms introduce one new functionally independent term. Additionally, since the derivatives are mixed between those seen in \eqref{longderiv1} and \eqref{quadderiv1}, we see that all isotropy will be fixed out at first order. The second order derivatives are
 
\begin{equation}
\begin{aligned}
&\begin{aligned}
    C_{kmkm;ll}&=C_{kmkl;lm}=C_{kmkl;ml}=C_{km\overbar{m}l;mm}=C_{km\overbar{m}l;\overbar{m}\overbar{m}}=C_{kl\overbar{m}l;k\overbar{m}}=C_{kl\overbar{m}l;\overbar{m}k}=C_{\overbar{m}l\overbar{m}l;kk}\\
    &=12K_0^2\left(\eta+\xi\right)^3\left(-g_3-g_2\xi+4\xi^3\right),
\end{aligned}\\  
&\begin{aligned}
C_{kmkm;\overbar{m}\overbar{m}}&=C_{kmkl;k\overbar{m}}=C_{kmkl;\overbar{m}k}=C_{km\overbar{m}l;kk}=C_{km\overbar{m}l;ll}=C_{kl\overbar{m}l;lm}=C_{kl\overbar{m}l;ml}=C_{\overbar{m}l\overbar{m}l;mm}\\
    &=12K_0^2\left(\eta+\xi\right)^3\left(g_3-g_2\eta+4\eta^3\right),
\end{aligned}\\
&\begin{aligned}
   C_{kl\overbar{m}l;m\overbar{m}}&=C_{kl\overbar{m}l;\overbar{m}m}=C_{\overbar{m}l\overbar{m}l;km}=C_{\overbar{m}l\overbar{m}l;mk}=C_{kl\overbar{m}l;lk}=C_{kl\overbar{m}l;kl}=C_{km\overbar{m}l;\overbar{m}l}=C_{km\overbar{m}l;l\overbar{m}}\\
   &=-C_{kmkm;l\overbar{m}}=-C_{kmkm;\overbar{m}\overbar{m}}=-C_{kmkl;kl}=-C_{kmkl;lk}\\
   &=-C_{km\overbar{m}l;km}=-C_{kmkl;\overbar{m}m}=-C_{km\overbar{m}l;mk}=-C_{kmkl;m\overbar{m}}\\
   &=\frac{3}{2}C_{kmkl;ll}=\frac{3}{2}C_{km\overbar{m}l;lm}=\frac{3}{2}C_{km\overbar{m}l;ml}=\frac{3}{2}C_{kl\overbar{m}l;mm}\\
   &=-\frac{3}{2}C_{kmkl;\overbar{m}\overbar{m}}=-\frac{3}{2}C_{km\overbar{m}l;k\overbar{m}}=-\frac{3}{2}C_{km\overbar{m}l;\overbar{m}k}=-\frac{3}{2}C_{kl\overbar{m}l;kk}\\
   &=12K_0^2\left(\eta+\xi\right)^3\sqrt{\left(g_3-g_2\eta+4\eta^3\right)\left(-g_3-g_2\xi+4\xi^3\right)},\\
\end{aligned}\\
&\begin{aligned}
    C_{kmkl;l\overbar{m}}&=C_{km\overbar{m}l;m\overbar{m}}=C_{km\overbar{m}l;\overbar{m}m}=C_{kl\overbar{m}l;km}\\
    &=3K_0^2\left(\eta+\xi\right)^3\left(-6g_3+3g_2\eta-3g_2\xi-16\eta^3+12\eta\xi^2+20\xi^3\right),\\
\end{aligned}\\
&\begin{aligned}
    C_{kmkl;\overbar{m}l}&=C_{km\overbar{m}l;kl}=C_{km\overbar{m}l;lk}=C_{kl\overbar{m}l;mk}\\
    &=3K_0^2\left(\eta+\xi\right)^3\left(-6g_3+3g_2\eta-3g_2\xi+16\xi^3-12\xi\eta^2-20\eta^3\right),
\end{aligned}    
\end{aligned}    
\end{equation}

\noindent which introduce no new functionally independent terms, and, thus, the algorithm stops here.

The non-vanishing CM invariants are
\begin{equation}
    w_1=-\frac{1}{\sqrt{6}}w_3^{2/3}=24K_0^2\left(\eta+\xi\right)^6,
\end{equation}
\noindent and the non-vanishing $\mathcal{I}$ invariants are
\begin{equation}
\begin{gathered}
       I_1=192K_0\left(\eta+\xi\right), \quad I_3=-2880K_0^3\left(\eta+\xi\right)^7\left(g_2-4\left(\eta^2-\eta\xi+\xi^2\right)\right), \\ I_5=-1327104K_0^5\left(\eta+\xi\right)^{13}\left(g_2-4\left(\eta^2-\eta\xi+\xi^2\right)\right).
\end{gathered}       
\end{equation}

 Note that there are cases where the $\mathcal{I}$ invariants will vanish and detect some special surface, although the physical meaning of these surfaces has yet to be understood.
\section{Conclusions}

In this paper, we discussed a number of historical solutions originally presented by Tullio Levi-Civita, outlining these solutions in a modern fashion. We also presented two different invariant characterizations of these solutions, using both the CK algorithm and CM invariants. As these methods are fully coordinate invariant they present a unique method of identifying these solutions for future work. 

The CK algorithm is used here to generate invariantly defined frames, along with curvature components projected along those frames. These invariant quantities are used to classify these spacetimes and provide insight into certain special cases which can now be seen to be identical. The CK invariants are of particular interest, as this method will extend to higher dimensional generalizations of these solutions in a natural way \cite{olver1995equivalence,mcnutt20175dcartan}. This paper should also serve as a useful reference for those who want to learn to use this algorithm in their own work, as the numerous cases presented here allow for one to see how the algorithm is applied under a number of different circumstances. 

The CM invariants, and in some cases $\mathcal{I}$ invariants, for these solutions are also constructed. These invariants provide a coframe independent classification of these solutions (to zeroth order), and also offer a useful characterization in cases where the CK invariants become complicated. In particular, the vanishing of certain invariants can be used to distinguish solutions that might otherwise appear deceivingly similar. The $\mathcal{I}$ invariants are also of interest, as there are cases in which these invariants vanish for certain coordinate values. This indicates that there may be interesting, and invariantly defined, surfaces present in these solutions which have not yet been analyzed.

As gravitational wave astronomy moves into the limelight of modern physics, solutions to Einstein's General Theory of Relativity have become increasingly important as a field of study even as an example for proof-by-contradiction. The high non-linearity of Einstein's field equations impose difficulties on finding solutions and an even greater difficulty in the interpretation of them. It is possible that invariantly analyzing these solutions could prove useful in understanding gravitational wave signals from novel sources. Though the 1917 Levi-Civita solution was not a wormhole, cylindrical symmetry seems to be a way to avoid topological censorship and consequently give hope to obtaining phantom-free wormholes, asymptotically flat in the radial direction \cite{wangcylindrical}. Of interest is the possibility that Levi-Civita’s metric could provide the pathway towards the engineering of artificial gravitational fields for human spaceflight.

\vspace{6pt}








\end{document}